\documentclass[10pt, conference]{IEEEtran}

\usepackage[margin = 0.7in]{geometry}
\usepackage{amsmath}
\usepackage{algpseudocode}
\usepackage{amssymb}
\usepackage{amsfonts}
\usepackage{amscd}
\usepackage{mathrsfs}
\usepackage{graphicx}
\usepackage[usenames,dvipsnames]{color}
\usepackage{bm}
\usepackage{url}
\usepackage{verbatim}
\usepackage[mathcal]{euscript}

\newtheorem{proposition}{Proposition}

\begin{document}

\title{Cognitive Radio from Hell: Flipping Attack on Direct-Sequence Spread Spectrum}

\author{J. Harshan$^{\dagger}$ and Yih-Chun Hu$^{\ast}$\\
$^{\dagger}$Indian Institute of Technology Delhi, India, $^{\ast}$University of Illinois Urbana-Champaign, USA\\
Email: jharshan@ee.iitd.ac.in, yihchun@illinois.edu\\}

\maketitle

%
%

\begin{abstract}
In this paper, we introduce a strong adversarial attack, referred to as the flipping attack, on Direct-Sequence Spread Spectrum (DSSS) systems. In this attack, the attacker, which is appropriately positioned between the transmitter and the receiver, instantaneously flips the transmitted symbols \emph{in the air} at 50\% rate, thereby driving the channel capacity to zero. Unlike the traditional jamming attack, this attack, when perfectly executed, cannot be detected at the receiver using signal-to-noise-ratio measurements. However, this attack necessitates the attacker to perfectly know the realizations of all the channels in the model. We first introduce the consequences of the flipping attack on narrowband frequency-flat channels, and subsequently discuss its feasibility in wideband frequency-selective channels. From the legitimate users' perspective, we present a method to detect this attack and also propose heuristics to improve the error-performance under the attack. We emphasize that future cyber-physical systems that employ DSSS should design transceivers to detect the proposed flipping attack, and then apply appropriate countermeasures.
\end{abstract}

\section{Introduction}
\label{sec:intro}

With wireless communication being an integral part of most cyber-physical systems \cite{Morris}, e.g. urban transportation, smart-grid and other IOT systems, it is imperative to not just secure wireless links from external attacks such as jamming, but also envision new attacks and provide suitable countermeasures against them. In this paper, we are interested in external attacks that can drive the channel capacity of communication between two legitimate users to zero. Although, jamming is a straightforward way of realizing such an objective, such attacks can be detected at the legitimate users by measuring their signal-to-interference-noise-ratio (SINR). In other words, jamming is not a stealth attack. From the jammer's perspective, while attack detection is a disadvantage, the jammer need not know the wireless channel between the source and the destination (other than the band of communication). In this paper, we would like to ask a converse question: \emph{Given perfect knowledge of the wireless channel between the source and the destination, is it possible for a sophisticated external attacker to drive their channel capacity to zero in stealth?}

To answer the above question, we envisage sophisticated threat models arising out of full-duplex radios \cite{DuSa} that operate as hidden relays in between a source and a destination. Loosely speaking, this threat comes under the well-known framework of \emph{man-in-the-middle attacks}, wherein the attacker can manipulate the transmitted symbols before they reach the destination. Although instantaneous modification of transmitted symbols has been addressed in theory to mitigate interference in wireless networks \cite{HJo, WDZLQL, GaHa}, this has not been studied as a threat to wireless security. In this work, we propose a new adversarial attack on wireless communication between two legitimate users, wherein the attacker, who is appropriately positioned between the two users, manipulates the symbols \emph{in the air}. Specifically, in the case of Binary Phase Shift Keying (BPSK), the attacker flips the BPSK symbols at 50\% rate independently, thereby driving the mutual information of the channel to zero. We refer to such an attacker as Cognitive Radio from Hell (CRFH), wherein the phrase \emph{from hell} highlights the legitimate users' inability to avoid this attack. The proposed attack from CRFH can be viewed as a form of \emph{correlated jamming} \cite{CJ1}, \cite{CJ2}, wherein the jammer has full or partial knowledge about the transmitter's signals. 

We first apply the proposed attack on frequency-flat narrowband communication channels, and subsequently discuss its impact on frequency-selective wideband communication channels. Due to practical constraints on applying this attack on wideband channels, we discuss a variant of the attack wherein the transmitted symbols on the delayed paths are manipulated, while keeping the symbol on the first significant path untouched. When perfectly executed, this attack can force the destination to combine the observations from all the paths, thereby degrading the error-performance. From the legitimate users' perspective, we discuss methods to detect this attack and also propose heuristics to improve the error-performance under the attack. Throughout the paper, we refer to the source, the destination, and the attacker as Alice, Bob, and Eve, respectively. 

\section{Flipping Attack on Narrowband Channels}
\label{sec:motivation}

\begin{figure}
\begin{center}
\includegraphics[scale=0.42]{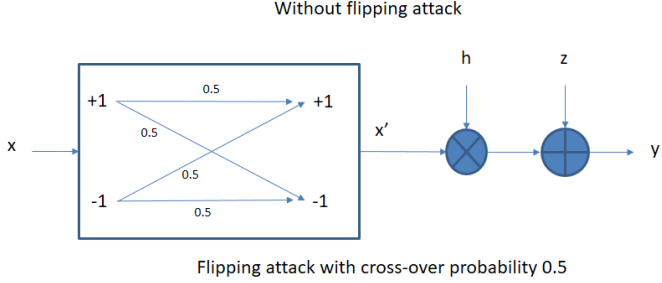}
\vspace{-0.7cm}
\caption{\label{fig:attack_block} In the flipping attack, the attacker instantaneously modifies the symbols in the air such that the receiver views the compound channel shown above. From information processing inequality, it is easy to verify that $I(x:y) = 0$ with perfect execution of the flipping attack. 
}
\end{center}
\end{figure}

Consider a narrowband communication channel between two legitimate players Alice and Bob (each equipped with single antenna), characterized by the signal model
\begin{equation}
\label{no_attack}
y_{k} = \sqrt{P}hx_{k} + n_{k},
\end{equation}
where $y_{k} \in \mathbb{C}$ is the received symbol by Bob at the $k$-th time-instant, $x_{k} \in \{-1, +1\}$ is the BPSK symbol transmitted by Alice, $n_{k} \in \mathbb{C}$ is the additive white Gaussian noise (AWGN) distributed as $\mathcal{CN}(0, \sigma^2)$, and $h \in \mathbb{C}$ represents the fading coefficient distributed as $\mathcal{CN}(0, 1)$. We have used binary phase shift keying (BPSK) constellations for the sake of introducing the flipping attack. However, this attack can also be generalized to higher-order constellations. The average signal-to-noise ratio (SNR) of this channel is $\frac{P}{\sigma^2}$. We assume a quasi-static fading channel where the realization $h$ is fixed over several consecutive symbol intervals. We denote the symbol period by $T_{s}$ seconds, and use $t_{main}$ to denote the time taken by the symbol to reach Bob. For the above described model, let us imagine a sophisticated adversarial attack initiated by Eve, who is physically positioned somewhere between Alice and Bob. We envisage a powerful attack, referred to as the flipping attack, wherein Eve is able to receive the transmitted symbol from Alice, decode it, and then transmit a suitably modified version to Bob within the symbol period. With such a strong attack model, let the additional processing-delay introduced by Eve be $t_{p}$ seconds, and the additional path-delay introduced by Eve be $t_{side}$ seconds. With that, the modified symbol reaches Bob after $t_{p} + t_{side}$ seconds. Provided we have 
\begin{equation}
\label{eq:flipping_condition}
t_{main} \leq t_{p} + t_{side} << t_{main} + T_{s},
\end{equation}
it is clear that Bob cannot resolve the transmitted symbol from Alice and the one manipulated by Eve. If we denote the manipulated symbol by $f(x, h, g)$, where $g$ is the channel between Eve and Bob, then the received symbol after the attack is 
\begin{equation}
\label{with_attack_general}
y_{k} = \sqrt{P} hx_{k} + g f(x_{k}, h, g) + n_{k}.
\end{equation}
In the flipping attack, Eve chooses the function $f(\cdot)$ such that $g f(x_{k}, h, g)  = -2\sqrt{P}hx_{k}$, which implies that Eve has perfect knowledge of $g$ and $h$. With such an ``in the air" modification, Bob will receive a flipped version of the transmitted symbol, given by
\begin{equation}
\label{with_attack_flip}
y_{k} = -\sqrt{P}hx_{k} + n_{k}.
\end{equation}
In the case of no attack, the received symbol is as given in \eqref{no_attack}. Note that if the attacker decides to flip the symbols at 50\% rate independently, then the BPSK symbols would go though the effective channel as shown in Fig.  \ref{fig:attack_block}. We can envision the attacker to flip the BPSK symbols by tossing a fair coin independently, thereby resulting in Bernoulli distribution with probability $0.5$. The following proposition on the above attack is straightforward to prove.

\begin{proposition}
From the information processing inequality \cite{CoT}, the mutual information $I(x:y)$ of the compound channel shown in Fig. \ref{fig:attack_block} is zero. 
\end{proposition}

The condition in \eqref{eq:flipping_condition} indicates that both path-delay and processing-delay through Eve are bottlenecks for perfectly executing the flipping attack. In extreme narrow-band communication, i.e., when the symbol period $T_{s}$ is much larger than the delay introduced by Eve, a perfectly executed flipping attack can drive the channel capacity to zero. However, in wideband communication, i.e., when $T_{s}$ is extremely small relative to the effective delay introduced by Eve, the modified symbol $g f(x, h, g)$ is likely to reach Bob after the current symbol period. This implies that Bob will have to treat the delayed modified symbol from Eve as noise, which in turn will lower the signal-to-interference-noise ratio (SINR) of the next transmitted symbol. In such a case, although the delayed modified symbol degrades the error-performance, this is not the intended consequence of the flipping attack. Recall that the primary objective of the flipping attack is to drive the channel capacity to zero in stealth.

\section{Flipping Attack on Wideband Channels}
\label{sec2:CRFH_wideband}

It is clear that driving the channel capacity to zero through the flipping attack is challenging when the symbol period $T_{s}$ is extremely small. However, a variant of this attack can be envisioned on a certain class of frequency-selective wideband channels, wherein multiple copies of the transmitted symbol arrive at Bob at different symbol periods; This way Eve gets sufficient time to execute the flipping attack on the delayed copies. Such channels are characterized by the signal model
\begin{equation}
\label{with_attack}
y_{k} = \sum_{l = 0}^{L_{d}-1} h_{l}x_{k - \tau(l)} + n_{k},
\end{equation}
where $y_{k}$ denotes the baseband sample received by Bob at $k$-th symbol period, and $\{h_{l} ~|~ 0 \leq l \leq L_{d}-1\}$ denotes the set of fading coefficients associated with the delayed copies. Here $\tau(l)$ denotes the delay of $l$-th copy measured in terms of number of samples. We refer to each of these copies as a \emph{tap}. Unlike in \eqref{no_attack}, we use $P = 1$ in the signal model in \eqref{with_attack}. The practical constraints on executing the flipping attack forbids Eve from flipping the BPSK symbol arriving on the first significant tap, i.e., on $h_{0}$. However, it is reasonable to assume that Eve can flip the BPSK symbols arriving on subsequent taps as she gets relatively longer time for processing and forwarding the received signals. In this case, we assume that Eve has perfect knowledge of the power-delay profile (PDP) of Alice-Bob's channel, and also its realizations.

Henceforth, throughout the paper, (i) $h_{0}$ is referred to as the main tap, whereas $\{h_{l} ~|~ 1 \leq l \leq L_{d} - 1\}$ are referred to as secondary taps, and (ii) flipping attack refers to the case when the symbol on the main tap is undisturbed, whereas the symbols on the secondary taps may be flipped independently at 50\% rate. In practice, the feasibility of executing the flipping attack depends on PDP of Alice-Bob's channel, particularly the delay of the secondary taps with respect to the main tap. Note that the objective of the attacker is to make sure that signals received on the secondary taps carry no information about the transmitted symbol. 


\subsection{Flipping Attack}

We assume that Alice and Bob communicate over a wideband channel using a DSSS system, wherein an $N$-length spreading code, which is securely shared between them, is applied on each of the BPSK symbols. We assume that Eve can accurately flip the chips with perfect knowledge of the channel estimates of Alice-Bob, Alice-Eve, and Eve-Bob. With perfect attack, Eve flips the BPSK symbols at 50\% rate. Note that once Eve decides to flip a BPSK symbol, she has to flip all the $N$ chips associated with that symbol. At the destination, Bob uses RAKE receivers to resolve the symbols arriving on the $L_{d}$ taps. After suitable correlation operation using the $N$-length spreading code on the received samples, Bob arrives at the $L_{d}$ equivalent received symbols, given by
\begin{equation*}
y_{k, l} = h_{l}x_{k} + z_{k, l}, ~0 \leq l \leq L_{d}-1,
\end{equation*}
where the new subscript $l$ is used to represent symbols received from the $l$-th finger, and $z_{k, l}$ is the effective AWGN, distributed as $\mathcal{CN}(0, \sigma^2)$, resulting from the correlation operation of the RAKE receiver. With that, the received symbols from the $L_{d}$ fingers are of the form
\begin{eqnarray*}
y_{k, 0} & = & h_{0}x_{k} + z_{k, 0},\nonumber \\
y_{k, l} & = & b_{k, l} h_{l}x_{k} + z_{k, l}, ~1 \leq l \leq L_{d}-1,
\end{eqnarray*}
where the polarity of $b_{k, l} \in \{-1, 1\}$ depends on attacker's choice. With uncoded communication, the Maximum Likelihood (ML) decoder expression is given by
\begin{equation}
\label{ML_decoding_combine_all_taps}
\tilde{x}_{k} = \arg min_{x \in \{-1, +1\}} \sum_{l = 0}^{L_{d}-1} |y_{k, l} - h_{l}x|^2,
\end{equation}%
where $h_{l}$ is the perfect estimate of the channel on the $l$-th tap. It is clear from \eqref{ML_decoding_combine_all_taps} that that attacker can force degraded error-performance by flipping symbols on some of the taps. On the defensive side, the receiver Bob may choose to use only the main tap fearing that the secondary taps might have been flipped. In such a case, the decoding operation is given by
\begin{equation}
\label{ML_decoding_first_tap}
\tilde{x}_{k} = \arg min_{x \in \{-1, +1\}} |y_{k, 0} - h_{0}x|^2.
\end{equation}
Note that although the attacker's signals are not affecting the error-performance using \eqref{ML_decoding_first_tap}, the overall error-performance will be worse than the no-attack case because Bob has not incorporated all the independent taps. In the next section, through simulations, we demonstrate the impact of the flipping attack on wideband frequency-selective channels with two taps and four taps in the delay-spread domain.

\subsection{Simulation Results}

In our setup, data communication between Alice and Bob takes place through a sequence of frames. Each frame constitutes $100$ BPSK symbols, out of which $20$ are occupied by the pilots. The pilot symbols, which are a priori fixed between Alice and Bob, also take values from BPSK constellation $\{-1, 1\}$. At Alice, a block of input bits of length 3968 bits are fed to a Rate-$\frac{1}{2}$ turbo encoder in feed-forward configuration [7 5], whose details are available in reference \cite{Turbo_code}. The corresponding output bits (7889 bits in number) are spread across the data part of several frames. Once the data and pilot symbols are packed, the frames are transmitted sequentially through a DSSS system, i.e., each BPSK symbol is multiplied by a spreading sequence of chip-rate $N = 128$. The frames are transmitted through the following wireless channels: (i) a two-tap channel with average power-profile $\{ E\{|h_{0}|^2\}, E\{|h_{1}|^2\}\} = \{0.5, 0.5\}$, and (ii) a four-tap channel with average power-profile $\{E\{|h_{0}|^2\}, E\{|h_{1}|^2\}, E\{|h_{2}|^2\}, E\{|h_{3}|^2\}\} = \{0.4, 0.3, 0.2, 0.1\}$. We assume that each $h_{l}$ is a circularly-symmetric complex Gaussian random variable, whose realization remains fixed throughout the frame, and takes an independent value in the next frame. 

Meanwhile, Bob uses RAKE receivers to resolve the dominant multipaths in the channel, and also estimates the channel realization on each tap using the pilot symbols. Subsequently, Bob combines the received symbols from all the taps to obtain log-likelihood ratio (LLR) on each BPSK symbol. Finally, the LLRs from each frame are forwarded to the turbo decoder, which processes them to decode the information bits. A total of $10$ iterations is used for the message passing algorithm in the turbo decoder. 

\begin{figure}
\begin{center}
\includegraphics[scale=0.36]{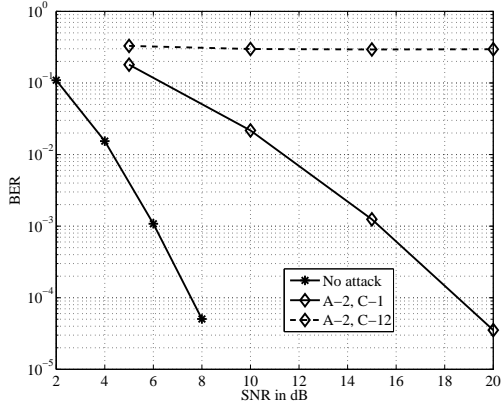}
\vspace{-0.5cm}
\caption{\label{fig:flipping_attack_2_taps}Turbo-coded error-performance on a two-tap channel with perfect estimate of Alice-Bob's channels at Eve. ``A-2, C-1" implies that Eve flips the symbols on the second tap at 50\% rate, while Bob combines the received symbols on first tap to obtain the LLR. Similarly, ``A-2, C-12" implies that Eve flips the symbols on the second tap at 50\% rate, while Bob combines the received symbols on the first and the second taps to obtain the LLRs. The plots show that combining the flipped symbols degrades the error-performance.}
\end{center}
\end{figure}

In Fig. \ref{fig:flipping_attack_2_taps}, we plot the Bit Error Rate (BER) performance of the above discussed system on the two-tap channel in three scenarios: (i) without attack - Eve does not flip the symbols on any tap, and Bob combines the observations on both taps, (ii) with attack - Eve flips the symbols on the second tap at 50\% rate independently, and Bob combines the observations on both taps, and finally (iii) with attack - Eve flips the symbols on the second tap at 50\% rate independently, and Bob discards the symbols received on the second tap. The plots indicate that an attack-ignorant Bob will experience error-floor behaviour in BER by blindly combining the observations on both taps, whereas an attack-aware Bob can recover the bits with some SNR loss by discarding the observations from the secondary tap. Similar experiments are also repeated for the four-tap channel, and the corresponding BER results are presented in Fig. \ref{fig:flipping_attack_4_taps}. In this case, we consider the following attack scenarios: (i) only the fourth tap is attacked, (ii) both the third and the fourth taps are attacked, (iii) second, third, and fourth taps are attacked. The plots in Fig. \ref{fig:flipping_attack_4_taps} show that Bob can avoid degraded error-performance if he can somehow detect the attacked taps and discard them when computing the LLR values. 

To obtain the simulation results in Fig. \ref{fig:flipping_attack_2_taps} and Fig. \ref{fig:flipping_attack_4_taps}, we have assumed that Eve knows the locations of the pilot symbols, and therefore she does not flip the pilots. As a result, Bob is able to estimate the channel on each tap accurately. However, since Eve strategically flips only the data symbols at 50\% rate, Bob is forced to combine all the taps as he is attack-ignorant. Thus, an important question to answer along this direction is \emph{how can Bob identify an unreliable tap?} This question will be addressed in the next section.

\subsection{Attack Detection Techniques}

From Eve's perspective, a critical task is to attack only on the data symbols. When this is ensured, Bob cannot detect the attack. On the other hand, when Eve flips the pilot symbols as well, then Bob can detect this attack by observing the polarity of the received symbols on the pilot locations. Therefore, from the legitimate users' perspective, in order to detect the flipping attack they must obfuscate the location of pilot symbols in every frame so that Eve is forced to flip some of the pilot symbols. To achieve this, we enable Alice and Bob to share a secret key using which the random positions of the pilots are determined. Since the positions of pilots are randomly chosen based on a pseudo-random generator, Eve cannot distinguish the pilots in the frame. Meanwhile, Bob's strategy is to observe the polarity of the received symbols on the pilot locations, and then detect the attack if at least one of the pilot symbols is flipped. 

By observing the polarity of the received symbols on the pilot locations, there is a non-zero probability with which Bob fails to detect the attack. At high SNR values, this event corresponds to the case when the positions of the flipped symbols do not coincide with the positions of the pilot symbols. 
At low SNR values, the change in the polarity of the pilot symbols may happen either due to the attack or due to the additive noise on each symbol. Therefore, to compute the probability of misdetection, we need to consider the event when the ambient noise unflips the pilot symbols already flipped by Eve. At arbitrary SNR values, the probability of misdetection for a given frame can be computed as
\begin{equation}
\label{eq:miss}
P_{miss}^{(l)} = \sum_{j = 1}^{L} p^{j}(1-p)^{L-j} \left[\sum_{x = 0}^{j} {L_{p} \choose x} {L - L_{p} \choose j - x} q_{l}^{x}\right],
\end{equation}
where ${n \choose r}$ denotes ``$n$ choose $r$" operation, and $q_{l} = \mbox{prob}(y_{k, l} < 0 ~|~ x_{k} = 1, h_{l})$, which is identical to $\mbox{prob}(y_{k, l} > 0 ~|~ x_{k} = -1, h_{l})$. To compute the expression in \eqref{eq:miss}, we assume that Eve flips the bits based on tossing a coin independently $L$ times. For a given frame, since the fading coefficient $h_{l}$ is constant, the value of $q_{l}$ is determined by the channel realization as well as the additive noise variance. 

\begin{figure}
\begin{center}
\includegraphics[scale=0.36]{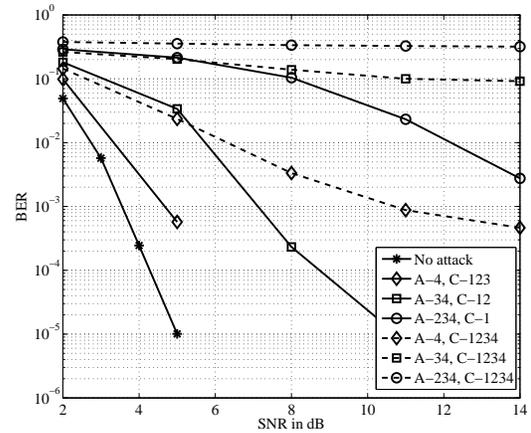}
\vspace{-0.5cm}
\caption{\label{fig:flipping_attack_4_taps}Turbo-coded error-performance on a four-tap channel with perfect estimate of Alice-Bob's channel at Eve. Notations in the legend are similar to those in Fig. \ref{fig:flipping_attack_2_taps}. The plots show that Eve has to flip the secondary taps with significant energy to force degraded error-performance at Bob.}
\end{center}
\end{figure}

\begin{figure}
\begin{center}
\includegraphics[scale=0.36]{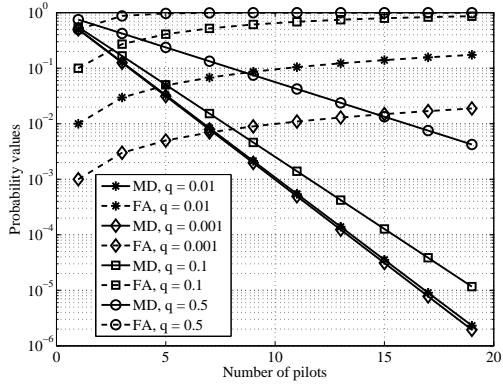}
\vspace{-0.3cm}
\caption{\label{fig:pd_pfa_flipping_attack}Probability of misdetection (MD) and probability of false alarm (FA) for various values of $L_{p}$ and $q$ (we discard the subscript $l$) with $L = 100$. The receiver detects the flipping attack if at least one of the received symbols on the pilot locations has changed its polarity.
}
\end{center}
\end{figure}

Similar to the events causing misdetection, events causing false alarm occur when at least one of the received symbols on the pilot locations is flipped due to the additive noise in the case of no attack. The corresponding probability of false alarm can be computed as
\begin{equation}
\label{eq:false_alarm}
P_{false}^{(l)} = 1 - (1-q_{l})^{L_{p}}.
\end{equation}
For a given SNR value, i.e., for a given $q_{l}$, the expression in \eqref{eq:false_alarm} indicates that $P_{false}^{(l)}$ increases as $L_{p}$ increases. However, the behavior of $P_{miss}^{(l)}$, given in \eqref{eq:miss}, as a function of $q$ is not straightforward. As a result, in the rest of this section, we plot $P_{miss}^{(l)}$ and $P_{false}^{(l)}$ against different values of $L_{p}$, $L$ and $q_{l}$. In Fig. \ref{fig:pd_pfa_flipping_attack}, we present the above values when $L = 100$ and when $L_{p}$ takes values from $\{1, 2, \ldots, 20\}$. The plots in Fig. \ref{fig:pd_pfa_flipping_attack} show that for a given value of $q_{l}$, $P_{miss}^{(l)}$ decreases as $L_{p}$ increases, while $P_{false}^{(l)}$ increases with $L_{p}$. However, when $q$ is sufficiently small (i.e., high SNR values), $P_{miss}^{(l)}$ can be reduced while keeping $P_{false}^{(l)}$ within acceptable range. This discussion shows that Bob can accurately detect the presence of Eve at high SNR values by observing the polarity of the received symbols on the pilot locations. Furthermore, with correct detection, Bob can decide whether to combine the received symbols on a secondary tap with the main tap or not. As shown in Fig. \ref{fig:flipping_attack_2_taps} and Fig. \ref{fig:flipping_attack_4_taps}, Bob can be conservative to drop the attacked taps, and just decode the information from the main tap only. In such a case, although there is error-performance loss, the receiver can still decode the information bits. On the other hand, if the attack-ignorant receiver combines all the taps without validating the polarity of the pilot symbols, then it would result in substantially degraded error-performance. 

In the next section, we consider the case when Eve does not have perfect knowledge of Alice-Bob's channel. With inaccurate estimate of the channel, we explore whether Bob can take advantage of this situation to improve the error-performance than that of just decoding from the main tap. 

\section{Flipping Attack: Incorrect Channel Estimates}
\label{sec:incorrect_estimates}

In practice, the knowledge of the channel estimate at Eve may not be accurate. Let the channel estimate of the $l$-th tap of Alice-Bob's channel at Eve be denoted by $\hat{h}_{l} = h_{l} + \epsilon_{l}$, where $h_{l}$ is the corresponding estimate at Bob and $\epsilon_{l}$ is the estimation error at Eve. Since we are focusing on one of the secondary taps, we henceforth drop the subscript $l$ in this section. With flipping attack using incorrect channel estimate at Eve, the received symbol at Bob is of the form $$y_{k} = (-h - 2\epsilon)x_{k} + z_{k}.$$ In the above expression, Eve has attempted to flip the transmitted symbol from $x_{k}$ to $-x_{k}$. However, because of incorrect channel estimate, the received symbol at Bob is offset by $-2\epsilon x_{k}$ compared to the case of perfect estimate. 

When the estimation error $\epsilon$ is non-negligible, we will show that Bob can identify the flipped data symbols, and subsequently undo the modifications to some extent. Let us assume a frame based communication between Alice and Bob with $L$ denoting the length of the frame and $L_{p}$ denoting the number of pilot symbols, which are transmitted at random positions in the frame. Since the positions of the pilots are generated based on a shared key, Eve does not know the pilot locations. Furthermore, since the pilots are also BPSK symbols, Bob can distinguish between a flipped and unflipped pilot symbol by observing the polarity of the received symbols. For the sake of exposition, we use an indicator variable $\mathcal{A}$ to represent the attack event. Within a frame, the expected value of the received symbols corresponding to the flipped BPSK symbols are
\begin{equation}
\label{avg_stat_attck_1}
E\{y_{k} | x_{k} = 1, \mathcal{A} = 1\} = -h -2\epsilon,
\end{equation}
\begin{equation}
\label{avg_stat_attck_2}
E\{y_{k} | x_{k} = -1, \mathcal{A} = 1\} = h + 2\epsilon,
\end{equation}
where the expectation $E\{\cdot\}$ is over the symbols of the frame. In \eqref{avg_stat_attck_1} and \eqref{avg_stat_attck_2}, $\mathcal{A} = 1$ indicates the attack event. In the case of no attack, similar values are given by
\begin{equation}
\label{avg_stat_no_attck_1}
E\{y_{k} | x_{k} = 1, \mathcal{A} = 0 \} = -h,
\end{equation}
\begin{equation}
\label{avg_stat_no_attck_2}
E\{y_{k} | x_{k} = -1, \mathcal{A} = 0\} = h.
\end{equation}
If $L_{p}$ is sufficiently large, Bob can obtain the estimates of the above statistics in \eqref{avg_stat_attck_1}-\eqref{avg_stat_no_attck_2} by observing the pilot symbols.\footnote{In order to obtain the statistics in \eqref{avg_stat_attck_1}-\eqref{avg_stat_no_attck_2}, we assume that Eve has flipped some pilot symbols from $-1$ to $1$, and some from $1$ to $-1$.} Then, if the difference $|E\{y_{k} | x_{k} = 1, \mathcal{A} = 1\} - E\{y_{k} | x_{k} = 1, \mathcal{A} = 0\}|$ is greater than $2\sigma$, then Bob proceeds to undo the flipping attack as discussed below. 

Using the above estimates, Bob will observe the rest of the $L - L_{p}$ symbols (the data symbols) in the frame, and then identifies the symbols that were flipped by Eve. The rationale behind this approach is that the flipped symbols are likely to be closer to  $-h -2\epsilon$ or $h +2\epsilon$, instead of $-h$ or $h$. Using this idea, Bob first identifies the locations of the data symbols that are closer to $-h -2\epsilon$ or $h +2\epsilon$ than $-h$ or $h$. Let these locations be denoted by $\mathcal{I} \subset\{1, 2, \ldots, L\}$. Similarly, Bob identifies the locations of the data symbols that are closer to $-h$ or $h$ than $-h -2\epsilon$ or $h +2\epsilon$. Let these locations be denoted by $\bar{\mathcal{I}}$. With this information, Bob changes the polarity of the received symbols on $\mathcal{I}$, while retaining the polarity of the symbols on $\bar{\mathcal{I}}$. Finally, the updated received symbols $\{y_{k}\}$ are suitably combined with those of the main tap in order to decode the information symbols. Specifically, for each symbol $y_{k}$, Bob combines it with the main tap depending on his confidence on how close the received symbol is to the offset versions. In particular, Bob generates an LLR as follows 
\begin{equation}
\label{eq:combine_taps_LLR}
\Delta_{k} = \mbox{log} \left(\frac{e^{\frac{-|y_{k} - (h+2\epsilon)|^2}{\sigma^{2}}} + e^{\frac{-|y_{k} - (-h-2\epsilon)|^2}{\sigma^{2}}}}{e^{\frac{-|y_{k} - h|^2}{\sigma^{2}}} + e^{\frac{-|y_{k} + h|^2}{\sigma^{2}}}}\right),
\end{equation}
which quantifies his confidence on whether the received symbol is close to $\{-h -2\epsilon, h +2\epsilon\}$ or $\{-h, h\}$. Here, $\sigma^{2}$ is the variance of the additive noise. From the above confidence metric, when $\Delta_{k} < -\delta_{th}$, for some optimization parameter $\delta_{th}$, Bob first changes the polarity of $y_{k}$ before including it with the corresponding symbol from the main tap. On the other hand, when $\Delta_{k} > \delta_{th}$, Bob uses $y_{k}$ as it is before including it with the corresponding symbol from the main tap. Finally and importantly, when $-\delta_{th} \leq \Delta_{k} \leq \delta_{th}$, Bob discards $y_{k}$, which implies that his confidence is not high to decide whether the symbol is flipped or otherwise. It is intuitive that when $\epsilon$ is small, Bob cannot confidently distinguish between flipped and unflipped data symbols, and therefore, the best strategy is to neglect those received symbols on the tap. Otherwise, combining the symbols despite low confidence would only degrade the error-performance as explained in the previous section. 

\begin{figure}
\begin{center}
\includegraphics[scale=0.34]{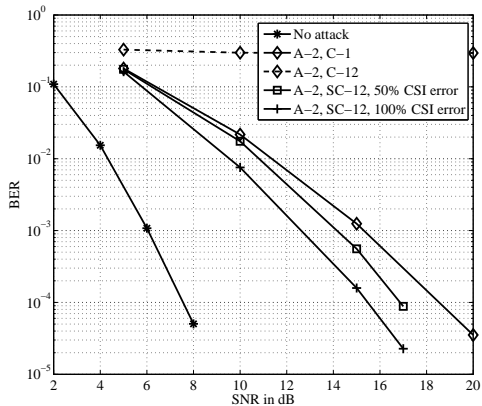}
\vspace{-0.5cm}
\caption{\label{fig:flipping_attack_2_taps_CE_error}Turbo-coded error-performance on a two-tap channel with imperfect estimate of Alice-Bob's channels at Eve. With channel estimation error at Eve, Bob opportunistically distinguishes between flipped and unflipped data symbols to some extent to improve the error-performance.  
}
\end{center}
\end{figure}

\begin{figure}
\begin{center}
\includegraphics[scale=0.32]{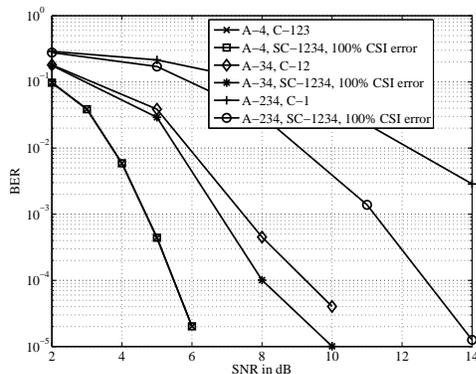}
\vspace{-0.5cm}
\caption{\label{fig:flipping_attack_4_taps_CE_error}Turbo-coded error-performance on a four-tap channel with imperfect estimate of Alice-Bob's channels at Eve.
}
\end{center}
\end{figure}

\subsection{Simulation Results}

In this section, we present simulation results to demonstrate that Bob can leverage on the estimation error at Eve to improve his error-performance compared to the conservative option of just using the main tap. The simulation setup in this section is same as in Section \ref{sec2:CRFH_wideband}. However, a major distinction is that the knowledge of the channel estimate at Eve is not accurate. We have chosen the estimation error to be either 50\% or 100\% to drive the point that Bob can significantly improve the performance. Bob first computes the estimates of the expressions in \eqref{avg_stat_attck_1}-\eqref{avg_stat_no_attck_2} using the pilot symbols, and then classifies the data symbols as either flipped or otherwise. The coded BER performance with incorrect estimate of the two-tap channel is presented in Fig. \ref{fig:flipping_attack_2_taps_CE_error}. In the presented results, ``A-2, C-1" denotes the case when the second tap is attacked, whereas Bob uses first tap to obtain the LLRs. Similarly, ``A-2, SC-12" denotes that second tap is attacked, and Bob smartly combines the symbols on the second tap with that of the first tap based on the LLR in \eqref{eq:combine_taps_LLR}. The plots in Fig. \ref{fig:flipping_attack_2_taps_CE_error} show that with larger values of $\epsilon$, Bob can opportunistically use the secondary tap to his advantage. We have used $\delta_{th} = 0.5$ as the threshold to distinguish the flipped and unflipped data symbols. 

We conducted similar experiments on the $4$-tap channel with average power-profile $\{0.4, 0.3, 0.2, 0.1\}$, and the corresponding results are presented in Fig. \ref{fig:flipping_attack_4_taps_CE_error}. We have assumed 100\% channel estimation error at Eve in this case. To obtain the results we used $\delta_{th} = 1$ for all the cases. Similar to the plots in Fig. \ref{fig:flipping_attack_2_taps_CE_error}, the plots in Fig. \ref{fig:flipping_attack_4_taps_CE_error} also show that the estimation error at Eve can be opportunistically used to improve the error-performance at Bob. It is interesting to note that the BER improvements from unflipping the symbols on the fourth tap (in the case of ``A-4, SC-1234") is negligible, whereas BER gains from unflipping the received symbols on the second, third and the fourth taps (in the case of ``A-234, SC-1234") are significant. This behaviour is attributed to the fact that the fourth tap alone contributes negligible signal power, whereas the signal power contributed by the second, the third and the fourth taps together are comparable to that of the main tap. We highlight that the choice of $\delta_{th}$ is crucial in reaping BER improvements from the attacked taps. While smaller values of $\delta_{th}$ include symbols on the unreliable taps into the decoding process thereby degrading the performance, larger values of $\delta_{th}$ forces Bob to discard the received symbols on the attacked taps, thereby matching the performance of that of combining the unattacked taps.

%

\section{Discussion and Directions for Future Work}
\label{sec:discussion}
We have discussed a strong adversarial attack on DSSS systems wherein the attacker instantaneously modifies the transmitted symbols such that some of the delayed copies carry no information on the transmitted data. Unlike jamming attack, this attack when perfectly executed, cannot be detected at Bob by measuring the SINR variations. Perfect execution of this attack necessitates the attacker to accurately know all the channel realizations in the model. Given that DSSS uses wideband communication, all the underlying channels in the model are frequency selective, and this implies that Bob may also receive multiple copies of the manipulated symbols transmitted by Eve. In this work, we have assumed that Eve nulls all the multipath components that she generates by converting the Eve-Bob's channel from frequency-selective to frequency-flat. However, in practice, this assumption needs unlimited power, and as a result, Bob may also receive multiple copies of the manipulated symbols from the attacker. It is interesting that Bob, who is oblivious to the presence of the attacker, may see more taps than that in Alice-Bob's channel, and the total number of taps depends on whether the delay profiles of Alice-Bob's and Eve-Bob's channels coincide. \emph{How can Bob opportunistically take advantage of no or imperfect nulling of multipaths in Eve-Bob's channel?} is an interesting direction for future work.

\section*{Acknowledgements}
This work was supported by the New Faculty Seed Grant - PLN06R to J. Harshan from the Indian Institute of Technology Delhi.

\end{document}